\documentstyle[12pt]{article}
\begin{document}

\author{N. Kiriushcheva and S.V. Kuzmin  \\ %EndAName
Department of Applied Mathematics\\University of Western Ontario, London,
N6A 5B7 Canada}
\title{Comments on ``The Einstein-Hilbert Lagrangian Density in a 2-dimensional
Spacetime is an Exact Differential'' by R. da Rocha and W.A. Rodrigues, Jr.}
\date{\today }
\maketitle

\begin{abstract}
We argue that the recent result of da Rocha and Rodrigues that in two
dimensional spacetime the Lagrangian of tetrad gravity is an exact
differential \cite{RR}, despite the claim of the authors, neither proves the
Jackiw conjecture \cite{Jackiw}, nor contradicts to the conclusion of \cite
{KK2}. This demonstrates that the tetrad formulation is different from the 
metric formulation of the Einstein-Hilbert action.
\end{abstract}

The Lagrangian density of two dimensional tetrad gravity has recently been
demonstrated, using the ``powerful and economic'' methods of Clifford
algebra by da Rocha and Rodrigues \cite{RR}, to be an exact differential.
Its explicit form is given in the last, fourteenth, section of \cite{RR}.
This well-known fact can hardly be called a new result 
(see, e.g., \cite{GKV}) and, in particular,
cannot be a good illustration of the power of the method used by authors. We
discuss the economy of this approach in the next section. The relevance of
this result \cite{RR} to the Jackiw conjecture \cite{Jackiw} and to the
conclusion of \cite{KK2} is the subject of the second section. In the third
section we comment on some particular results in \cite{RR}. The last section
is our conclusion.

{\bf 1.} We start by reviewing some results of the tetrad formulation, in
order to establish notation. Tetrads are a set of orthogonal vectors $e_\mu
^a$ erected at each point of $D$-dimensional spacetime ($\mu =0,...,D-1$ are
world coordinates indices and $a=0,...,D-1$ are tetrad indices). The
Lagrangian of tetrad gravity (TG) can be obtained by performing a direct
substitution of the metric tensor $g_{\mu \nu }$ in terms of tetrads, $%
g_{\mu \nu }=e_{a\mu }e_\nu ^a$, into the Einstein-Hilbert (EH) Lagrangian
(e.g., Eq. (9) of \cite{DeserNeiuw1974}). This can also be done by using
particular combinations built from tetrads and their derivatives, such as
Ricci rotational coefficients \cite{Eisenhart} ($\gamma _{abc}\equiv e_{a\mu
;\nu }e_b^\mu e_c^\nu $ which is a world scalar), spin connections 

\begin{equation}
\label{0}(\omega _{\mu ab}\equiv -e_{a\nu ;\mu }e_b^\nu =-e_{a\nu ,\mu
}e_b^\nu +\Gamma _{\nu \mu }^\lambda e_{a\lambda }e_b^\nu ,
\end{equation}
which is a world vector), etc.; and then using properties of the commutator
of covariant derivatives of a covariant vector in Riemannian space $%
V_{\alpha ;\beta ;\gamma }-V_{\alpha ;\gamma ;\beta }=V_\sigma R_{\;\alpha
\gamma \beta }^\sigma $ (see, e.g., p. 291 of \cite{LL} or p. 49 of \cite
{Wald}). We are then left with

\begin{equation}
\label{1}L\left( e\right) =eR\left( e\right) 
\end{equation}
with

$$
R\left( e\right) =R\left( g\left( e\right) \right) =R\left( g\left( e\right)
,\Gamma \left( e\right) \right) =R\left( e,\gamma \left( e\right) \right)
=R\left( e,\omega \left( e\right) \right)  
$$
where $e=\det \left| e_\mu ^a\right| $ and semicolons ``$;$'' indicate
covariant differentiation.

One particularly common form of the Lagrangian of tetrad gravity is (see,
e.g. \cite{Schwinger1963}; Eq. (11-12) of \cite{DeserNeiuw1974}; Eq. (2.1)
of \cite{CNP1982})\footnote{%
The different sign of (\ref{2}) in some articles (e.g., \cite{DeserNeiuw1974}%
) is, probably, the result of using different from \cite{LL}, \cite{Carmeli}%
, \cite{Wald} convention for Riemannian tensor as, for example, in \cite
{Weinberg}.}

\begin{equation}
\label{2}L\left( e\right) =ee^{a\mu }e^{b\nu }\left( \omega _{\nu ab,\mu
}-\omega _{\mu ab,\nu }+\omega _{\mu ac}\omega _{\nu \;b}^{\;c}-\omega _{\nu
ac}\omega _{\mu \;b}^{\;c}\right) .
\end{equation}

Using (\ref{0}) and the expression for $\Gamma _{\nu \mu }^\lambda \left(
g\left( e\right) \right) $ one easily obtains (Eq. (7) of \cite
{DeserNeiuw1974}; Eq. (2.4) of \cite{DeserIsham1976})

\begin{equation}
\label{3}\omega _{\nu ab}=\frac 12\left[ e_a^\rho \left( e_{b\rho ,\nu
}-e_{b\nu ,\rho }\right) -e_b^\rho \left( e_{a\rho ,\nu }-e_{a\nu ,\rho
}\right) +e_a^\rho e_b^\mu \left( e_{c\rho ,\mu }-e_{c\mu ,\rho }\right)
e_\nu ^c\right] . 
\end{equation}

In two dimensions ($2D$), because of the antisymmetry of $\omega $ in the
tetrad indices ($\omega _{\mu ab}=-\omega _{\mu ba}$) it immediately follows
that the part of (\ref{2}) that is quadratic in $\omega $ ,

\begin{equation}
\label{3a}L_{\omega \omega }=ee^{a\mu }e^{b\nu }\left( \omega _{\mu
ac}\omega _{\nu \;b}^{\;c}-\omega _{\nu ac}\omega _{\mu \;b}^{\;c}\right) ,
\end{equation}
is zero, and the only non-zero part of (\ref{2}) is 

\begin{equation}
\label{4}L\left( e\right) =2\omega _{0(0)(1),1}-2\omega _{1(0)(1),0}.
\end{equation}
(We use $()$ brackets to distinguish explicit values of tetrad indices.)
Equation (\ref{4}) follows from the antisymmetry of $\omega $ in the tetrad
indicies and the $2D$ relation 

\begin{equation}
\label{4a}e\left( e^{\left( 0\right) 0}e^{\left( 1\right) 1}-e^{\left(
1\right) 0}e^{\left( 0\right) 1}\right) =-1
\end{equation}
since $e=e_0^{\left( 0\right) }e_1^{\left( 1\right) }-e_0^{\left( 1\right)
}e_1^{\left( 0\right) }$ (we use $\eta _{ab}=\left( +-\right) $ for lowering
and rising tetrad indices). Equation (\ref{4}) can also be derived using
differential forms.

To express the $2D$ Lagrangian of TG as a total divergence in a manifestly
covariant (tensorial) form, we can use a trick similar to what was employed
in the metric formulation (see p. 269 \cite{LL} or p. 98 of \cite{Carmeli}
and for its use in the Hamiltonian formulation of the EH Lagrangian \cite
{Pirani}, \cite{Dirac1958}, \cite{KK2}) in order to express terms with
second order derivatives of the tetrads as a total derivative. This can be
done directly in terms of tetrads or, simplier, by working with $\omega
_{\mu ab}$.\footnote{%
Another similarity between $\Gamma $ and $\omega $ is the equivalence of the
first and second order formulations for the metric tensor and tetrad fields
in higher than two dimensions, when $g_{\mu \nu }$ and $\Gamma $, and,
respectively, $e_\mu ^a${\tt \ and }$\omega ${\tt ,} are treated as
independent fields. Note, that for both of them the first and second order
formulations are not equivalent in $2D$.} This converts (\ref{2}) into

\begin{equation}
\label{5}L\left( e\right) =-\left[ e\left( e^{a\mu }e^{b\nu }-e^{a\nu
}e^{b\mu }\right) \right] _{,\mu }\omega _{\nu ab}+V_{,\mu }^\mu +L_{\omega
\omega },
\end{equation}
where

\begin{equation}
\label{6}V^\mu =\left[ e\left( e^{a\mu }e^{b\nu }-e^{a\nu }e^{b\mu }\right)
\right] \omega _{\nu ab}=2ee^{a\mu }e^{b\nu }\omega _{\nu ab.} 
\end{equation}

In $2D$ $L_{\omega \omega }=0$ and the first term also vanishes because all
non-zero contributions to the first term, when written in terms of
components, give a constant value to the term in square brackets, as can be
seen from (\ref{4a}). Hence, in $2D$ the Lagrangian of TG is a total
derivative of a vector $V^\mu $ ($L_{2D}\left( e\right) =V_{,\mu }^\mu $),
and this expression is equivalent to (\ref{4}). In $2D$, $V^\mu $ can be
written in a very simple form

\begin{equation}
\label{6a}V^\mu =2\varepsilon ^{ab}\varepsilon ^{\nu \rho }e_a^\mu e_{b\rho
,\nu }
\end{equation}
where $\varepsilon $ is the totally antisymmetric tensor ($\varepsilon
^{01}=\varepsilon ^{(0)(1)}=1$). Note, that (\ref{6}) was obtained using just
one step by rearranging the terms in (\ref{2}) leading to (\ref{5}). One
more manipulation is needed; we must substitute (\ref{3}) into (\ref{6})
which in $2D$ gives (\ref{6a}). This is a compact form of the main result of 
\cite{RR} (see Eqs. (35-39) of section fourteen). This result also is given
by Eq. (1.55) of \cite{GKV}, where it is obtained using a different method.
Equation (1.55) of \cite{GKV} is equivalent to (\ref{6a}).

{\bf 2.} We will now show that 
this well-known result (derived using a longer method in \cite{RR})
neither proves the Jackiw conjecture \cite{Jackiw}, nor contradicts the
conclusion of \cite{KK2}.

The Jackiw conjecture (p. 353 of \cite{Jackiw}, Ref. 1 of \cite{RR}), is
that ``...{\it in two dimensions it} [Einstein theory] {\it cannot even be} 
{\it formulated since }$G_{\mu \nu }$ [Einstein tensor] v{\it anishes
identically. Correspondingly, the Hilbert-Einstein action is a surface term}%
...''. More generally, 
it appears that 
according to this conjecture, from triviality of the
equations of motion, it follows that the action is a surface term. This is
the converse of the statement that Lagrangians which are pure divergences
give trivial equations of motion.

This conjecture can be shown to be incorrect by finding a counter example,
while, at the same time, finding a single example (such as $2D$ tetrad
gravity) is not sufficient for establishing its proof. One particular
counter example is the Einstein-Hilbert Lagrangian density when expressed in
terms of the metric. This observation was made by Deser and Jackiw (see the
second sentence after Eq. (2.8) on p. 1504 of \cite{Deser-Jackiw} (Ref. 2 of 
\cite{RR})): ``${\it R}^\mu $ [such as $\partial _\mu R^\mu =\sqrt{-g}R$] 
{\it cannot be presented explicitly and locally in terms of the generic
metric }$g_{\mu \nu }${\it \ and its derivatives }$\partial _\alpha g_{\mu
\nu }$''. See also p.18 of Strobl \cite{Strobol} ``..{\it .it does not seem
to be possible to express }$\sqrt{-g}R${\it \ explicitly as a covariant
total differential of }$g${\it \ itself''}. 
An explicit demonstration of this is provided in \cite{KK2}. 
How can this be reconciled with
the results for $2D$ tetrad gravity?

The result of \cite{KK2} was obtained for the EH (metric) Lagrangian and
confirmed the above statements. This was used to find a Hamiltonian
formulation of non-divergent part of the metric Lagrangian in a way similar
to the treatment \cite{Pirani} of the second order form of the EH Lagrangian
in higher dimensions. In $2D$, use of the ADM formulation \cite{ADM} leads
to the unphysical result that there are negative degrees of freedom \cite
{Martinec}; i.e. it is an overconstrainted system. In contrast, 
the results of \cite{KK2} are completely self-consistent.

It appears that the source of confusion about apparent contradiction between
the results contained in \cite{KK2} (so as the statements quoted above) and
the well-known result for tetrad gravity (rederived in \cite{RR}) is based
on the commonly held belief that the EH Lagrangian, when expressed in terms
of the metric and in terms of the tetrad, are equivalent, and, moreover,
since the two Lagrangians equivalent, then if one of them is a total
divergence the second one also must be a total divergence (this would be the
logical conclussion if the Jackiw conjecture could be proven). 
It is important to note though that in \cite{RR}
it is not established that the two forms of the EH action are equivalent,
and hence we cannot conclude that since the tetrad form of this Lagrangian
is a total divergence, then so is the metric form.

We now will consider the question of whether the two Lagrangians are really
equivalent. 

Even though both Lagrangians lead to the trivial equations of motion we
cannot call these Lagrangians equivalent. The part of the EH Lagrangian
(when expressed in terms of the metric) that is not a surface term (the $%
\Gamma \Gamma $-part only) is not generally covariant \cite{Pirani} and is
not a true scalar \cite{Carmeli}. Though the equations of motion are the
same both with and without the surface term, the part of Lagrangian without
a surface term is not generally covariant. To consider question of 
equivalence, then all terms have to be taken into account since in the
presence of a surface contribution, we cannot rely only on the 
equations of motion beeing equivalent. 
The role of a surface term can be analyzed using a 
generalization of the Hamiltonian procedure (e.g., see \cite{Regge}).
Recently, the peculiar features of surface terms has been reconsidered
from quite a different point of view in \cite{Pad}.

Further, it was stated a long time ago by Einstein in his first article on
tetrad ($n$-bein) gravity \cite{Einstein1928}: ``{\it The }${\it n}${\it %
-bein field is determined by }$n^2${\it \ functions }$e_a^\mu $ [tetrads], 
{\it whereas the Riemannian metric is determined just by }$\frac{n\left(
n+1\right) }2${\it \ quantities. According to (3)} [$g_{\mu \nu }=e_{a\mu
}e_\nu ^a$]{\it , the metric is determined by the }${\it n}${\it -bein field
but not vice versa}''. This inequivalence was his main reason for
introducing tetrads; by modifying the EH Lagrangian a unification of gravity
with electrodynamics might be possible. Different models considered include
also modifications of Riemannian space itself to Riemann-Cartan, Weyl, etc.
spaces. A list of the different spaces examined is contained in Eqs.
(4.103)-(4.107) on p. 105 of \cite{Waldyr}.

One can find different equivalent formulations of the EH action by going
from one set of field variables (fields) to another, provided the functional
Jacobian of such a transformation is non-singular. For example, we recently
used the particular linear combinations $\xi _{\alpha \beta }^\lambda
=\Gamma _{\alpha \beta }^\lambda -\frac 12\left( \delta _\alpha ^\lambda
\Gamma _{\beta \sigma }^\sigma +\delta _\beta ^\lambda \Gamma _{\alpha
\sigma }^\sigma \right) $ in Hamiltonian formulation of the first order EH
action beyond two dimensions \cite{KK1}, \cite{KKM3}. The Jacobian of the
transformation between $\Gamma $ and $\xi $ is non-singular and field
independent. Elimination of the fields $\xi $ using their equations of
motion leads back to the EH action, providing a proof of the equivalence of
the first and second order formalisms. However, for a singular change of
field variables (it is not, probably, correct even to call this ``a change
of variables'') without being able to invert it ({\it ``not vice versa''}),
we instead generate a new model which, though it might be an interesting, is
not equivalent to the original one, at least mathematically. We obviously
are dealing with a singular case when the number of fields in the two sets
of variables is different (e.g., $n^2$ of tetrad components and $\frac{%
n\left( n+1\right) }2$ components of the metric tensor).

Moreover, even if these (EH and TG) mathematically non-equivalent
Lagrangians give physically equivalent results (such as, for example, the
same number of degrees of freedom, the same observables, etc.) it would not
be enough to draw the conclusion that if the $2D$ tetrad Lagrangian is a
total divergence, then the EH (metric) Lagrangian is also a total divergence.

Some authors define physical equivalence in very broad ways. For example,
according to \cite{Kaempffer}: {\it ``All experimental knowledge about
gravitation existing at present is therefore compatible with any theory that
coincides for weak fields with the linearized version of Einstein's theory.
In particular, a vierbein field theory of gravitation is sufficiently
supported by experiment, provided the field equations written in terms of
the variables $e_\alpha ^k$, defined by }$e_\alpha ^k=\delta _\alpha ^k+\eta
_\alpha ^k$,{\it \ reproduce in an approximation linear in these variables
the physical content of Einstein's linearized theory''}. This definition of
equivalence is hard to accept.

We note that the tetrad formulation was introduced to couple fermionic
fields to gravity because, as was shown by Cartan, one cannot couple them
directly to the metric (see \cite{DeserNeiuw1974}, \cite{DeserIsham1976} and
references therein). Another motivation to use tetrads is to have more
similarities with ordinary gauge theories than the metric formulation has 
\cite{UtiyamaKibble}. Can these facts serve as an additional indication
that, at least, in some sense the EH and TG Lagrangians are different?

{\bf 3.} Now we would like to provide some additional and less general
comments on particular results and conclusions (statements) of \cite{RR}.

3.1. According to the authors of \cite{RR} (Sect. 8), their Eq. (17) ``{\it %
...are first order Lagrangian densities (first introduced by Einstein)...}''. 

First of all, we would like to mention that though Einstein, and not
Palatini \cite{Pal} (asis generally believed), was the first who introduced
first order formalism using the affine-metric formulation \cite{Einstein1925}%
, he never discussed the first order formulation of tetrad gravity
(Riemannian space), Eq. (17) of \cite{RR}, and was interested only in the
theory of distant teleparallelism (which is an example of non-Riemannian space,
see, e.g., Definition 236 of \cite{Waldyr}, Ref. 9 of \cite{RR}).\footnote{%
Einstein used this formulation in 1928-1930 papers in his second version of
the unified theory. In 1931 he wrote \cite{Science} ``...we reached the
conclusion that we were striving in the wrong direction...''. He abandoned
this approach and never returned to it again.} 

Secondly, the first order tetrad-spin connection formulation in $2D$ is not
equivalent to the second order (tetrad) formulation \cite{LR}. This is
because the equation of motion for spin connection in $2D$ does not result
in (\ref{3}). This is similar to the EH Lagrangian, where in $2D$ the first
(affine-metric) and second order (metric) formulations are also not
equivalent \cite{Mann}, \cite{LR} as the equation of motion for the affine
connection does not yield the Christoffel symbol (App. A of \cite{KK1}).
Consequently, in $2D$ $\omega $ must be considered to be a dependent field,
denoting the function of tetrads given by (\ref{3}). 

3.2. The sentences in the end of Sect. 8 and the beginning of Sect. 9 of 
\cite{RR} seem to contradict each other: ``{\it ...the statement that in
general coordinate chart }$L_g\neq 0${\it \ in a 2-dimensional spacetime is
correct...}'' and in almost the next line ``{\it ...let us first show that }$%
L_g=0${\it \ in a 2-dimensional spacetime}'' and further ``{\it which
implies also that corresponding }$L_{\Gamma \Gamma }=0$''. Let us comment on
the last conjecture that from $L_g=0$ follows $L_{\Gamma \Gamma }=0$. 

We note that, in addition to the fact that the metric and tetrad formulations
are different and do not have a very simple correspondence with each other, 
it is also true
for PARTS of these different Lagrangians. In particular, the part $L_g^o$
(Eq. (22)) and Eq. (23) of \cite{RR} are not equivalent. One way to
demonstrate this is to use the definition of $\omega $ in (\ref{0}). Solving
(\ref{0}) for $\Gamma $, we see that

\begin{equation}
\label{7}\Gamma _{\beta \mu }^\alpha =\omega _{\mu ab}e^{a\alpha }e_\beta
^b+e_{a\beta ,\mu }e^{a\alpha },
\end{equation}
and then substituting this into the $\Gamma \Gamma $-part of the EH action

\begin{equation}
\label{8}L_{\Gamma \Gamma }=\sqrt{-g}g^{\alpha \beta }\left( \Gamma _{\sigma
\lambda }^\lambda \Gamma _{\alpha \beta }^\sigma -\Gamma _{\sigma \beta
}^\lambda \Gamma _{\alpha \lambda }^\sigma \right) ,
\end{equation}
one obtains

\begin{equation}
\label{9}L_{\Gamma \Gamma }\left( g\left( e\right) ,\Gamma \left( e,\omega
\right) \right) =L_{\omega \omega }+L_{extra}
\end{equation}
with

\begin{equation}
\label{10}L_{extra}=ee^{f\alpha }e_f^\beta \left( e_{,[\beta }^{a\lambda
}e_{a\alpha ,\lambda ]}+e^{a\lambda }\omega _{[\lambda ab}e_{\alpha ,\beta
]}^b-e_\alpha ^b\omega _{[\beta ab}e_{,\lambda ]}^{a\lambda }\right) ,
\end{equation}
where square brackets indicate antisymmetrization $h_{...[\beta
...}^{...}g_{...\lambda ]...}^{...}=h_{...\beta ...}^{...}g_{...\lambda
...}^{...}-h_{...\lambda ...}^{...}g_{...\beta ...}^{...}$.

This manipulation is valid in any dimension but, what is more interesting is
the fact that $L_{extra}\neq 0$ even in $2D$, making the conjecture that $%
L_{\Gamma \Gamma }=L_{\omega \omega }$ in \cite{RR} to be incorrect. $%
L_{extra}$ is absent in the total Lagrangian of TG because the substitution
of (\ref{7}) into the part of the EH Lagrangian containing derivatives of $%
\Gamma $ gives

\begin{equation}
\label{11}\sqrt{-g}g^{\mu \nu }\left( \Gamma _{\mu \nu ,\lambda }^\lambda
-\Gamma _{\mu \lambda ,\nu }^\lambda \right) =ee^{a\mu }e^{b\nu }\left(
\omega _{\nu ab,\mu }-\omega _{\mu ab,\nu }\right) -L_{extra}.
\end{equation}
so, in the total Lagrangian $L_{extra}$ cancels.

3.3. Equation (29) (Sect. 11 of \cite{RR}) is a total differential for the
tetrad gravity, not for the EH Lagrangian (metric), and authors confirmed 
that ``%
{\it ...(29) NEEDS the introduction of tetrad field to be written}''. Does
it contradict the result of \cite{KK2} and the above mentioned statements
of Deser and Jackiw \cite{Deser-Jackiw} and Strobl \cite{Strobol}? Is it the
proof that the EH (metric) Lagrangian is an exact differential in $2D$, as
stated in the title of \cite{RR}, if one finds the introduction 
of tetrads unavoidable to write the final result?

3.4. In Sect. 14 of \cite{RR} the ``divergence term'' is calculated (Eq.
(35-39)) by using the ``powerful and economic'' formalism. The resulting
vector is again expressed in terms of tetrads, so, as in our previous
comment, the introduction of tetrads is NEEDED again. Moreover, for the
tetrad gravity the same result can be easily obtained directly (as we
demonstrated, see Eq. (\ref{6})).\footnote{%
What can, probably, illustrate the power of the method used in \cite{RR} (or
authors) is the ability in two days to obtain two absolutely different
solutions (see V.1 on December 14 and V.2 on December 16 of \cite{RR}).
However, the method of \cite{RR} does not seem very powerful compare with
methods of the string theory that in twenty years reach much larger number ($%
10^{500}$) of possible solutions \cite{Nature} that makes it the theory of
``more than everything''.} The necessity of using tetrads to constract an
expression of a vector $V^\mu $ such that $L_{2D}\left( e\right) =V_{,\mu
}^\mu $ was emphasized before in Ref. 11 of \cite{Deser-Jackiw}, where it
was also stated that this vector {\it ``depends on the essentially
non-metric part''. }Indeed,{\it \ }part of it cannot be expressed in terms
of a metric field and hence, it is not possible to construct such a vector for
the $2D$ EH (metric) Lagrangian.

3.5. The last paragraph in the end of Sect. 14 (which is a kind of
``Conclusion'') is even difficult to call a conjecture. Here any reader of 
\cite{RR} can only guess about the relation of the $2D$ tetrad gravity to
the Hamiltonian formulation of the EH Lagrangian considered in \cite{KK2},
and about even more mysterious connections with different $2D$ MODELS of 
\cite{KKM1}, \cite{KKM2} (Refs. 5,6 of \cite{RR}) and, probably, with all
the existing (e.g., \cite{GKV}) and $2D$ models not yet devised. Even if,
according to the authors of \cite{RR}, there is no consistent Hamiltonian
formulation for pure divergent Lagrangians, it is not clear why $2D$ models 
\cite{GKV}, \cite{KKM1}, \cite{KKM2} with non-divergent Lagrangians also do
not have a consistent Hamiltonian formulation.

The consistency of the Hamiltonian formulation is based on the closure of
the Dirac procedure \cite{Dirac}, \cite{Sund}, \cite{GT} (closure of the
constraint algebra) as well as possibility of finding gauge transfromations
from a knowledge of the first class constraints. These transformations can be
verified directly to be a symmetry of the Lagrangian. Such transformations
leave the Lagrangian invariant, and follow from the consistent Hamiltonian
treatment of the EH action in $2D$. 

It is not clear how the fact, that the TG Lagrangian is a total
divergence in $2D$ implies that the Hamiltonian formulation
of different $2D$ models is inconsistent as is claimed in \cite{RR}. 
Possibly the authors of \cite{RR} found some
inconsistencies in the Hamiltonian formulation of $2D$ tetrad gravity,
and again extrapolated this result onto the EH Lagrangian and all $2D$ models. 
In principle, one can try to find a Hamiltonian formulation even for models 
(such as $%
2D$ tetrad gravity) which are total divergences but, for example, consist of
terms linear in second order derivatives and quadratic in first order
derivatives. In such cases, one can try to treat these Lagrangians as higher
order derivative theories, and apply the appropriate formulation for such 
models, due to Ostrogradsky \cite{Ostr}, or its modification \cite{GT}
for the singular (gauge invariant) cases.\footnote{%
The first attempt to consider Hamiltonian formulation of EH Lagrangian as
higher order theory with a purpose to preserve general covariance is due to
Dutt and Dresden \cite{DD}.} For the $2D$ EH (metric) Lagrangian, the
application of generalized Ostrogradsky method gives a consistent
Hamiltonian formulation \cite{KK3}, which supports the result obtained in 
\cite{KK2}.

{\bf 4. }In conclusion we would like to emphasize that in order to discuss
how the Einstein-Hilbert (metric) and the tetrad gravity Lagrangians are
equivalent, one has to take into account the effect of all terms in these
Lagrangians, including surface terms, because all terms are needed to
preserve general covariance. In dimensions higher than two, we can avoid
``the surface term problem'' by using first order formulation, but this is
not possible in $2D$. However, generalization of the Hamiltonian formulation
for higher order derivative Lagrangians or some other modifications (e.g.,
see p. 54 of \cite{Thiemann} and references therein) may provide a means of
doing so. Dirac himself conjectured \cite{Dirac}, {\it ``I} [Dirac] {\it feel
that there will always be something missing from them} [non-Hamiltonian
methods] {\it which we can only get by working from a Hamiltonian''.}

We also note that this equivalence cannot be discussed at the expense of
giving up general covariance. For example, if in $2D$ we set $e_0^{\left(
1\right) }=e_1^{\left( 0\right) }=0$, so as $g_{01}=0$, and consider only
subset of coordinate transformations that preserve this initial
restrictions, these two formulations may be even become equivalent, because
when $g_{01}=0$ the Einstein-Hilbert (metric) Lagrangian in fact becomes a
total divergence \cite{KK2} and the number of independent fields (those which 
are
left) are equal for both Lagrangians ($g_{00}$ and $g_{11}$ for the metric, $%
e_0^{\left( 0\right) }$ and $e_1^{\left( 1\right) }$ for the tetrads). There
is one-to-one correspondence among them: $g_{00}=\left( e_0^{\left( 0\right)
}\right) ^2$, $g_{11}=-\left( e_1^{\left( 1\right) }\right) ^2$ and $%
L_{extra}$ (\ref{10}) is zero. However, true equivalence between the tetrad
and metric formulations cannot be established only by fixing some field
variables and imposing a restriction on the general coordinate transformations.

We would like also to add that the question of equivalence between different
formulations of gravity is not purely a $2D$ problem, but it is also
related, for example, to the new variables \cite{Ashtekar} which, according
to \cite{Henneaux}, are equivalent to tetrads. These new variables are
exclusively used in attempts to quantize GR and, in particular, they are
variables of loop quantum gravity (for a recent review see \cite{Nicolai})
which is now the best developed \cite{Smolin} proposal for the theory of
quantum gravity. It is not evident that a formulation of gravity in terms of
these variables is equivalent to a formulation in terms of the metric.\\

{\bf Acknowledgment}\\

The authors are grateful to D.G.C. McKeon for discussions and reading the
manuscript. We thank S.B. Gryb and R. Nowbakht Ghalati for discussions and
continuous interest to our work.

Our special thanks to D. Grumiller who emphasized (before appearance of \cite
{RR}) the necessity to explain the difference of results for metric and
tetrad gravity. We also thank J.S. Kuzmina for pointing out on \cite{Nature}.

\end{document}